\documentclass[pra,aps,twocolumn,showpacs,superscriptaddress]{revtex4}
\usepackage{graphics}
\usepackage{amsmath}
\begin{document}
\title{Robustness of non-Gaussian  entanglement against noisy
  amplifier and attenuator environments}
 \author{Krishna Kumar Sabapathy}
 \email{kkumar@imsc.res.in}
 \affiliation{Optics and Quantum Information Group, The Institute of Mathematical
   Sciences, Tharamani, Chennai 600 113, India.}
 \author{J. Solomon Ivan}
 \affiliation{Raman Research Institute, C. V. Raman Avenue, Sadashivanagar, 
 Bangalore 560 080, India.}
 \author{R. Simon}
 \affiliation{Optics and Quantum Information Group, The Institute of Mathematical
   Sciences, Tharamani, Chennai 600 113, India.}
\begin{abstract}
The recently developed Kraus representation for bosonic Gaussian
channels is employed to study analytically the robustness of
non-Gaussian entanglement against evolution under noisy attenuator and
amplifier environments, and compare it with the robustness of Gaussian
entanglement. Our results show that some non-Gaussian
states with one ebit of entanglement are more robust than all Gaussian 
states, even the ones with arbitrarily large entanglement, a 
conclusion of
direct consequence to the recent conjecture by Allegra
et.\,al.\,[PRL, {\bf 105}, 100503 (2010)]. 
\end{abstract}
\pacs{03.67.Mn, 03.65.Yz, 42.50.Dv, 03.67.-a}
\maketitle

%

Early developments in quantum information technology of continuous 
variable (CV) systems largely concentrated on Gaussian states and 
Gaussian operations\,\cite{sam-rmp}. The symplectic group of linear 
canonical transformations\,\cite{dutta94} is available as a handy and 
powerful tool in this Gaussian scenario, leading to an elegant
classification of permissible Gaussian processes or 
channels\,\cite{caruso}. 
The fact that states in the non-Gaussian sector could offer advantage 
for several quantum information tasks has resulted more recently in 
considerable interest in non-Gaussian 
states, both experimental\,\cite{exp} and 
theoretical\,\cite{theory}.

Since  noise is unavoidable in any actual realization of these 
information processes\,\cite{kimble}, robustness of entanglement and other 
nonclassical effects against noise becomes an important consideration. 
 Allegra et al\,\cite{allegra} have thus studied the evolution of 
what they call {\em photon number entangled states} (PNES) (i.e., 
two-mode states of the form $|\psi\rangle = \sum\,c_n\,|n,\,n\rangle$) 
in a  {\em noisy} attenuator environment. They conjectured based on  
numerical evidence that, for a given energy, Gaussian entanglement is 
more robust than the non-Gaussian ones. Earlier Agarwal et 
al\,\cite{agarwal091} had shown that entanglement of the NOON state is 
more robust than Gaussian entanglement in the {\em quantum limited} 
amplifier environment. More recently, Nha et al\,\cite{carmichael10} have 
shown that nonclassical features, including entanglement, of several 
non-Gaussian states survive a {\em quantum limited} amplifier 
environment much longer than Gaussian entanglement. Since the 
 conjecture of Ref.\,\cite{allegra} refers to the noisy environment  
 and  the analysis in Ref.\,\cite{agarwal091,carmichael10} to the noiseless or 
quantum-limited case, the conclusions of the latter do not necessarily
amount to 
refutation of the conjecture of Ref.\,\cite{allegra}. Indeed, Adesso has 
argued very recently\,\cite{adesso} that the well known 
extremality\,\cite{extremality} of Gaussian states implies 
proof and rigorous validation of the conjecture of 
Ref.\,\cite{allegra}.     

In the present work we employ the recently developed\,\cite{kraus10} 
Kraus representation of bosonic Gaussian channels to study 
analytically the behaviour of non-Gaussian states in {\em noisy} 
attenuator and amplifier environments. Both NOON states and a simple 
form of PNES are considered. Our results show conclusively that the 
conjecture of Ref.\,\cite{allegra} is too strong to be maintainable.

{\em Noisy attenuator environment}\,: Under evolution through a noisy
attenuator channel ${\cal 
  C}_{1}(\kappa,a), \, \kappa \leq 1$, an input state ${\rho}^{\rm in}$
with characteristic 
function (CF) $\chi_{W}^{\rm{in}}(\xi)$ goes to state ${\rho}^{\rm
  out}$ with CF 
\begin{eqnarray}
\chi^{\rm{out}}_{W}(\xi)= \chi^{\rm{in}}_{W}(\kappa \xi)
 \,e^{-\frac{1}{2} (1- \kappa^2 
  + a )|\xi|^2},
\label{eq1}
\end{eqnarray}
where $\kappa$ is the attenuation 
parameter\,\cite{caruso}. 
 In this notation, 
quantum
limited channels\,\cite{carmichael10} correspond to $a=0$, and so the 
parameter 
$a$ stands
for the {\em additional Gaussian noise}. 
Thus, ${\rho}^{\rm in}$ is taken under the
two-sided symmetric action of  ${\cal C}_{1}(\kappa,
a)$ to ${\rho}^{\rm out}= {\cal C}_{1}(\kappa, a) \otimes {\cal
  C}_{1}(\kappa, a) \, ({\rho}^{\rm in})$
with CF 
\begin{align}
\chi^{\rm{out}}_{W}(\xi_1, \xi_2) = 
\chi^{\rm{in}}_{W}(\kappa \xi_1, \kappa \xi_2)
 \,e^{-\frac{1}{2} (1- \kappa^2 +a)(|\xi_1|^2 + |\xi_2|^2)}.
\label{eq2}
\end{align}
To test for separability of ${\rho}^{\rm out}$ we may
 implement the partial transpose test on ${\rho}^{\rm
  out}$ in the Fock basis or on
$\chi^{\rm{out}}_{W}(\xi_1, \xi_2)$. {\em The choice could depend on 
the state}.

Consider first the Gaussian case, and in particular 
the two-mode squeezed state
$|\psi(\mu)\rangle = {\rm sech}\mu \sum_{n=0}^{\infty} \tanh^n \mu
|n,n \rangle$ with variance matrix $V_{\rm sq}(\mu)$. 
Under the
two-sided action of noisy attenuator channels ${\cal C}_{1}(\kappa,
a)$, 
 the output two-mode Gaussian state ${\rho}^{\rm{out}}(\mu) 
= {\cal C}_{1}(\kappa, a) \otimes {\cal
  C}_{1}(\kappa, a) \, (\,|\psi(\mu)\rangle \langle\psi(\mu)|\,)$
has variance matrix  
\begin{eqnarray}  
V^{\rm out}(\mu) &=& \kappa^2 V_{\rm sq}(\mu) + (1-\kappa^2 +
a){1\!\!1}_{4}, \,\,\, \nonumber \\
V_{\rm sq}(\mu)&=& \left(\begin{array}{cc}
c_{2\mu}{1\!\!1}_{2} & s_{2\mu}\sigma_3 \\
s_{2 \mu}\sigma_3 & c_{2 \mu} {1\!\!1}_{2}
\end{array} \right), 
\label{eq3}
\end{eqnarray}
where 
 $c_{2\mu}= \cosh 2\mu$, $s_{2 \mu}= \sinh 2\mu$. 
Note that our variance matrix differs from that
of some authors by a factor 2; {\em in particular, the
variance matrix of  vacuum is the unit matrix in our notation}. 
 Partial transpose test\,\cite{simon00} shows that
${\rho}^{\rm out}(\mu)$ is separable iff $a \geq
\kappa^2 (1 - e^{-2 \mu})$. The `additional noise' $a$
required to render ${\rho}^{\rm out}(\mu)$ separable is an 
increasing function of the squeeze (entanglement) parameter $\mu$ 
and saturates at $\kappa^2$. 
In particular, $|\psi(\mu_1)
\rangle$, $\mu_1 \approx 0.5185$ corresponding to one ebit of
entanglement 
is rendered separable when $a \geq \kappa^2 (1- e^{-2 \mu_1})$.
 For $a \geq \kappa^2$, ${\rho}^{\rm out}(\mu)$ is separable,
independent of the initial squeeze parameter $\mu$. {\em Thus  
$a = 
\kappa^2$
is the additional noise that renders separable all Gaussian states.}

Behaviour of non-Gaussian entanglement may be handled directly in the 
Fock basis using 
 the recently developed Kraus
representation of Gaussian channels\,\cite{kraus10}.
  In this basis 
quantum-limited attenuator ${\cal C}_1(\kappa;0), \, \kappa \leq 1$ 
and 
quantum-limited amplifier ${\cal C}_2(\kappa;0), \, \kappa \geq 1$ 
are described, respectively, by Kraus operators\,\cite{kraus10}
\begin{eqnarray*}
B_{\ell}(\kappa) &=& \sum_{m=0}^{\infty}
\sqrt{{}^{m+\ell} C_{\ell}}\, 
(\sqrt{1-\kappa^2}\,)^{\,\ell}\, {\kappa}^{m}  
| m \rangle \langle m+\ell|,\nonumber\\ 
A_{\ell}(\kappa) &=& \frac{1}{\kappa} \sum_{m=0}^{\infty} 
\sqrt{{}^{m+\ell} 
  C_{\ell}} 
(\sqrt{1-\kappa^{-2}}\,)^{\,\ell}\frac{1}{\kappa^m}
| m +\ell \rangle \langle m |,
\end{eqnarray*}
$\ell =0,1,2,\cdots$. In either case, the noisy channel ${\cal
  C}_j(\kappa;a), \, j=1,2$ can be 
realized in the form ${\cal C}_2(\kappa_2;0) \circ {\cal
  C}_1(\kappa_1;0)$, so that the Kraus operators
for the noisy case is simply  the product set
$\{A_{\ell^{\,'}}(\kappa_2) B_{\ell}(\kappa_1)  \} $. 
Indeed, the
composition rule ${\cal C}_2(\kappa_2;0) \circ {\cal
  C}_1(\kappa_1;0) = {\cal C}_1(\kappa_2\kappa_1;2(\kappa_2^2-1))$ 
or ${\cal C}_2(\kappa_2\kappa_1;2\kappa_2^2(1-\kappa_1^2))$ 
according as $\kappa_2\kappa_1 \leq 1$ or $\kappa_2\kappa_1 \geq 1$
implies that the noisy attenuator ${\cal C}_1(\kappa;a), 
\kappa \leq 1$ is realised by the choice $\kappa_2 = \sqrt{1 + a/2}
\geq 1, \, \kappa_1 = \kappa/\kappa_2 \leq \kappa \leq 1$, and the noisy
amplifier 
${\cal  C}_2(\kappa;a), \, \kappa \geq 1 $ by 
$\kappa_2 = \sqrt{\kappa^2 +
  a/2} \geq \kappa \geq 1, \, \kappa_1 = \kappa/\kappa_2 
\leq 1$\,\cite{kraus10}.   
 {\em Note that one goes from realization of ${\cal 
C}_1(\kappa;a),\,\kappa\le 1$ to that of
  ${\cal C}_2(\kappa;a),\,\kappa\ge 1$ simply by replacing $(1+a/2)$ 
by 
$(\kappa^2 +
  a/2)$; this fact will be exploited later.}

Under the action of ${\cal C}_j(\kappa;a) = {\cal C}_2(\kappa_2;0) \circ
{\cal C}_1(\kappa_1;0),\,j=1,2$  the elementary operators 
$|m\rangle\langle n|$ 
go to 
\begin{align}
&{\cal C}_{2}(\kappa_2; 0) \circ {\cal C}_{1}(\kappa_1; 0) \left( |m
  \rangle \langle n| \right) \nonumber \\ &= 
\kappa_2^{-2} \sum_{j=0}^{\infty} \sum_{\ell=0}^{\text{min}(m,n)}
\left[ {}^{m-\ell+j}C_j\,
  {}^{n-\ell+j}C_j\,{}^mC_{\ell}\,{}^nC_{\ell}\right]^{1/2} 
\nonumber \\ 
&~~~\times (\kappa_2^{-1} \kappa_1)^{(m+n-2\ell)}\,  (1-\kappa_2^{-2})^j
\,(1-\kappa_1^2)^{\ell}  
\nonumber \\ 
&~~~\times |m-\ell +j \rangle \langle n-\ell +j|.
\label{nm}
\end{align}
Substitution of $\kappa_2 = \sqrt{1+a/2}, \kappa_1 = \kappa/\kappa_2$
gives realization of ${\cal C}_1(\kappa;a), \kappa \leq 1$ while
$\kappa_2 = \sqrt{\kappa^2+a/2}, \kappa_1 = \kappa/\kappa_2$ gives
 that of ${\cal C}_2(\kappa;a), \, \kappa \geq 1$. 

As our first non-Gaussian example we study the NOON state $| \psi
\rangle = \left(|n0\rangle + |0n \rangle  \right)/\sqrt{2} $ with 
density matrix 
\begin{align}
{\rho} &= \frac{1}{2} \left( |n\rangle \langle n| \otimes |0
  \rangle \langle 0| + |n\rangle \langle 0| \otimes |0
  \rangle \langle n| \right. \nonumber \\
& ~~~+|0\rangle \langle n| \otimes |n
  \rangle \langle 0| + |0\rangle \langle 0| \otimes |n
  \rangle \langle n| \left.  \right).
\end{align}
 The output state 
${\rho}^{\rm out} = {\cal C}_{1}(\kappa ; a) \otimes {\cal
  C}_{1}(\kappa ; a)( \rho)$ can be
detailed in the Fock basis through use of Eq.\,(\ref{nm}).

To test for inseparability, we
project ${\rho}^{\rm out}$ 
 onto the $2 \times 2$ subspace spanned by the four bipartite vectors
$\{|00\rangle,\, |0n \rangle, \, |n,0 \rangle, \, |n,n \rangle \}$, and
test for entanglement in this subspace;  this
simple test proves sufficient for our purpose! The matrix
elements of interest are\,:  
$\rho^{\rm out}_{00,00}$, $\rho^{\rm out}_{nn,nn}$, and $\rho^{\rm 
out}_{0n,n0} = \rho^{\rm
  out\,*}_{n0,0n}$.  Negativity of
$\delta_{1}(\kappa, a) \equiv {\rho}^{\rm out}_{00,00} {\rho}^{\rm
  out}_{nn, nn} - |{\rho}^{\rm out}_{0n, n0}|^2 $ will prove for 
$\rho^{\rm out}$  not only NPT entanglement, but
also one-copy distillability \cite{horodecki97}. 

\begin{figure}
\scalebox{1.1}{
\includegraphics{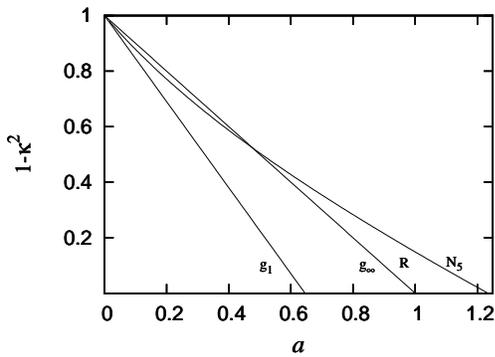}}
\caption{
Comparison of the robustness of the entanglement of a NOON
state with that of two-mode Gaussian states under the 
two-sided action of symmetric noisy attenuator. 
\label{fig1}} 
\end{figure}

To evaluate ${\rho}^{\rm out}_{00,00}$, ${\rho}^{\rm out}_{0n,n0}$, and
${\rho}^{\rm out}_{nn,nn}$, it suffices to
evolve the four single-mode operators $|0\rangle\langle0|$, $|0 \rangle
\langle n|$, $|n \rangle \langle 0|$, and $|n \rangle \langle n|$
through the noisy attenuator ${\cal C}_1(\kappa;a)$ using
Eq.\,\eqref{nm}, and then project the output to one of these
operators. For our purpose we need only the 
following single mode matrix elements\,:
\begin{eqnarray}
x_1 &\equiv& \langle n| {\cal C}_1(\kappa;a) (|n\rangle \langle n|)| n
\rangle \nonumber\\ 
&=&
(1+a/2)^{-1} \sum_{\ell=0}^{n} \left[ {}^n C_{\ell} \right]^2 [\kappa^2 
(1+a/2)^{-2}]^{\ell} 
\nonumber \\
& \times& [(1-\kappa^2(1+a/2)^{-1}) (1-(1+a/2)^{-1})]^{n -\ell}\, , \nonumber \\
x_2 &\equiv& \langle 0| {\cal C}_1(\kappa;a) (|n\rangle \langle n|)| 0
\rangle \nonumber \\
&=& (1+a/2)^{-1} [1-\kappa^2(1+a/2)^{-1}]^n, \nonumber \\
x_3 &\equiv& \langle 0| {\cal C}_1(\kappa;a) (|0\rangle \langle 0|)| 0 \rangle = (1+a/2)^{-1},\nonumber \\
x_4 &\equiv& \langle n| {\cal C}_1(\kappa;a) (|0\rangle \langle 0|)| n
\rangle \nonumber \\
&=&(1+a/2)^{-1}[1-(1+a/2)^{-1}]^n ,\nonumber \\
x_5 &\equiv& \langle n| {\cal C}_1(\kappa;a) (|n\rangle \langle 0|)| 0 \rangle =
\kappa^n (1+a/2)^{-(n+1)},\nonumber \\
 &\equiv& \langle 0| {\cal C}_1(\kappa;a) (|0\rangle \langle n|)| n
 \rangle^{*}  
\label{elements}
\end{eqnarray}
One finds ${\rho}^{\rm out}_{00,00}= x_2 x_3$, ${\rho}^{\rm
  out}_{nn,nn} = x_1 x_4$, and ${\rho}^{\rm out}_{0n,n0} = x^2_5/2$,
and therefore \begin{align} \delta_1(\kappa, a)= x_1x_2x_3 x_4 -
(|x_5|^2/2)^2, \end{align} 
Let $a_1(\kappa)$ be the solution to $\delta_1(\kappa,a)=0$. This
means that entanglement of our NOON state survives all values of noise
$a < a_1(\kappa)$. The curve labelled $N_5$ in Fig.\,\ref{fig1}
 shows, in the $(a,\,\kappa)$ space,  $a_1(\kappa)$ for
the NOON state with $n=5$\,:  entanglement of $(| 50 \rangle + |05
\rangle)/\sqrt{2} $ survives all noisy 
attenuators below $N_5$. The straight line denoted $g_{\infty}$
corresponds to $a = \kappa^2$\,: channels above this line break
entanglement of all Gaussian states, even the ones with arbitrarily
large entanglement. The line $g_1$ denotes $a
= \kappa^2(1- e^{-2\mu_1})$, where $\mu_1 = 0.5185$ 
 corresponds to 1 ebit of Gaussian entanglement\,:
Gaussian entanglement $\leq 1$ ebit does not survive any of the
channels above this line. The region $R$ of channels above $g_{\infty}$
but below $N_5$ are distinguished in this sense \,: {\em no Gaussian
entanglement survives the channels in this region, but the} NOON {\em state 
$(| 50
\rangle + | 0 5 \rangle)/\sqrt{2} $ does}.  

As a second non-Gaussian example we study the PNES 
$|\psi \rangle  = \left(|00 \rangle + | nn \rangle
\right)/\sqrt{2}$ with density matrix 
\begin{align}
{\rho} = &\frac{1}{2} \left( |0\rangle \langle 0| \otimes |0
  \rangle \langle 0| + |0\rangle \langle n| \otimes |0
  \rangle \langle n| \right. \nonumber \\
& ~~+|n\rangle \langle 0| \otimes |n
  \rangle \langle 0| + |n\rangle \langle n| \otimes |n
  \rangle \langle n| \left.  \right).
\end{align}
The output
state ${\rho}^{\rm out} = {\cal C}_{1}(\kappa;
a)\otimes {\cal C}_{1}(\kappa; a) \, \left( {\rho} \right)$ can be 
detailed in the Fock basis through use of Eq.\,(\ref{nm}).

Now to test for entanglement of ${\rho}^{\rm out}$, we project
again ${\rho}^{\rm out}$ onto the $2 \times 2$ subspace spanned by the 
vectors 
$\{ |00\rangle,\, |0n \rangle, \, |n,0 \rangle, \, |n,n \rangle\}$,
and see if it is (NPT) entangled in this subspace. Clearly, it suffices
to evaluate the matrix elements ${\rho}^{\rm 
  out}_{0n,0n}$, ${\rho}^{\rm out}_{n0,n0}$, and 
${\rho}^{\rm out}_{00,nn}$, for  if
$\delta_{2}(\kappa, a) \equiv {\rho}^{\rm
  out}_{0n,0n} {\rho}^{\rm out}_{n0,n0}-
|{\rho}^{\rm out}_{00,nn} |^2$
is negative  then ${\rho}^{\rm out}$ is NPT
entangled, and one-copy distillable. 

\begin{figure}
\scalebox{1.1}{
\includegraphics{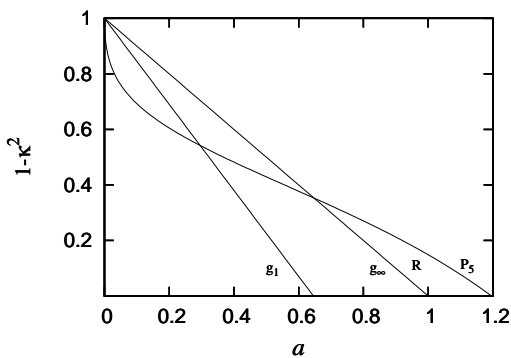}}
\caption{Comparison of the robustness of the entanglement of a PNES
state with that of two-mode Gaussian states under the action of
two-sided symmetric noisy attenuator. 
\label{fig2}} 
\end{figure}

Once again, the matrix elements listed in  \eqref{elements} prove 
sufficient to determine $\delta_2(\kappa,a)$:  ${\rho}^{\rm
  out}_{0n,n0} = {\rho}^{\rm out}_{n0,n0} = (x_1x_2 + x_3x_4)/2$,
and ${\rho}^{\rm out}_{00,nn} = |x_5|^2/2$, and so 
\begin{align}
\delta_2(\kappa,a) = ((x_1 x_2 + x_3x_4)/2)^2 - (|x_5|^2/2)^2. 
\end{align}
Let $a_2(\kappa)$ denote the solution to $\delta_2(\kappa,a) =0$. That
is, entanglement of our PNES survives all $a \leq
a_2(\kappa)$. 
 This $a_2(\kappa)$ is shown as the curve labelled $P_5$ in 
Fig.\,\ref{fig2} for 
the
PNES $(|00 \rangle + | 55 \rangle )/\sqrt{2}$. The lines $g_1$
and $g_{\infty}$ have the same meaning as in Fig.\,\ref{fig1}. The 
region $R$ above
$g_{\infty}$ but below $P_5$ corresponds to channels
$(\kappa,a)$ under whose action all two-mode Gaussian states are
rendered separable, while entanglement of the non-Gaussian PNES $(|00
\rangle + | 55 \rangle )/\sqrt{2}$ definitely survives.

{\em Noisy amplifier environment}\,: 
We turn our attention now to the amplifier environment. Under the
symmetric two-sided 
action of a noisy amplifier channel ${\cal C}_{2}(\kappa; 
a), \, \kappa \geq 1$, the  two-mode CF 
$\chi^{\rm{in}}_{W}(\xi_1,\xi_2)$ is  
taken to
\begin{eqnarray*}
\chi_W^{\rm{out}}(\xi_1,\xi_2) = \chi^{\rm{in}}_W(\kappa \xi_1, \kappa
\xi_2)\,e^{-\frac{1}{2}(\kappa^2-1+a)(|\xi_1|^2 + |\xi_2|^2)}. 
\label{tr2}
\end{eqnarray*}
In particular, the two-mode
squeezed vacuum state $|\psi(\mu) \rangle$ with variance matrix
$V_{\rm sq}(\mu)$ is taken to a Gaussian state with variance matrix
\begin{eqnarray}
V^{\text{out}}(\mu) =\kappa^2 V_{\text{sq}}(\mu) + 
(\kappa^2-1 + a) {1\!\!1}_{4}.
\label{eq10}
\end{eqnarray} 
The partial transpose test\,\cite{simon00} readily shows that 
the output
state is separable when $a \geq 2-\kappa^2 (1+e^{-2\mu})$: 
the additional noise $a$ required to render the output Gaussian state 
separable  increases with the squeeze or entanglement parameter $\mu$ 
and saturates at $a =  2- \kappa^2$: for $a \geq 2- \kappa^2$ 
the output state is separable
for every Gaussian input. The noise required to render 
the two-mode squeezed state $|\psi(\mu_1)
\rangle $ with 1 ebit of
entanglement 
($\mu_1\approx 0.5185$) separable is $a = 2-\kappa^2(1+e^{-2\mu_1}) $.
\begin{figure}
\scalebox{1.1}{
\includegraphics{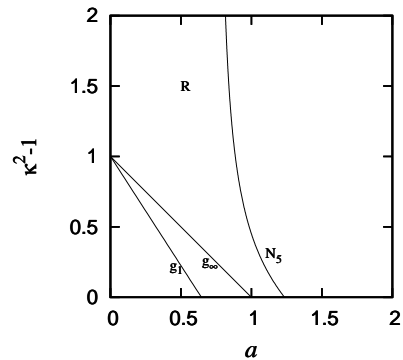}}
\caption{Comparison of the robustness of the entanglement of a NOON
state with that of all two-mode Gaussian states under the action of
two-sided symmetric noisy amplifier.
\label{fig3}} 
\end{figure}
Now we examine the behaviour of the NOON state $(|n0 \rangle + | 0n 
\rangle)/\sqrt{2}$
under the symmetric action of noisy amplifiers ${\cal
  C}_2(\kappa;a),\, \kappa \geq 1$.  Proceeding exactly
as in the attenuator case, we know that ${\rho}^{\rm{out}}$ is definitely
entangled if $\delta_3(\kappa,a) \equiv {\rho}^{\rm 
out}_{00,00} {\rho}^{\rm
  out}_{nn, nn} - |{\rho}^{\rm out}_{0n, n0}|^2  $ is negative. As
remarked earlier the expressions for ${\cal
   C}_1(\kappa;a), \, \kappa \leq 1$ in Eq.\,\eqref{elements} are 
valid 
for ${\cal
   C}_2(\kappa;a), \, \kappa \geq 1$ {\em {provided}} $1 + 
a/2$ is
replaced by $\kappa^2 + a/2$. For clarity we denote by $x^{\,'}_j$ the
expressions resulting from $x_j$ when ${\cal C}_1(\kappa;a), \, \kappa \leq 1$
replaced by ${\cal 
  C}_2(\kappa;a), \, \geq 1$ and $1 + a/2$ by $\kappa^2 + a/2$. For
instance, $x_5^{\,'} \equiv \langle n| {\cal C}_2(\kappa;a) (\,|n\rangle \langle 0|\,)| 0 \rangle =
\kappa^n (\kappa^2+a/2)^{-(n+1)} $ and $\delta_3(\kappa;a) =
x_1^{\,'} x_2^{\,'} x_3^{\,'} x_4^{\,'} - (|x_5^{\,'}|^2/2)^2$. 

Let $a_3(\kappa)$ be the solution to $\delta_3(\kappa,a)=0$. This is
represented  in Fig.\,\ref{fig3} by the curve marked $
N_5$, for the case of NOON state $(|05 \rangle + |50 \rangle
)/\sqrt{2}$. This curve is to be compared  with the line $a = 2 - \kappa^2$, denoted
$g_{\infty}$, above which no Gaussian entanglement survives, and with
the line $a = 2 - \kappa^2(1 + e^{-2\mu_1}), \, \mu_1= 0.5185$,
denoted $g_1$, above which no Gaussian entanglement $\leq 1$ ebit
survives. In particular, {\em the region $R$ between $g_{\infty}$ and 
$N_5$
corresponds to noisy amplifier channels against 
which entanglement of the} NOON {\em state $( |05 \rangle   + |50\rangle
)/\sqrt{2}$ is robust, whereas no Gaussian entanglement survives}. 

\begin{figure}
\scalebox{1.1}{
\includegraphics{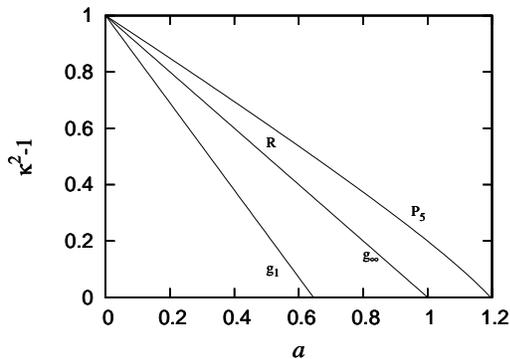}}
\caption{
Comparison of the robustness of the entanglement of a PNES
state with that of all two-mode Gaussian states under the action of 
 two-sided symmetric noisy amplifier.  
\label{fig4}} 
\end{figure}

Finally, we consider the behaviour of the PNES $( |00 \rangle + |nn 
\rangle )/\sqrt{2}$ in this noisy amplifier environment. The
output, denoted ${\rho}^{\rm{out}}$, is certainly entangled if
$\delta_4(\kappa,a) \equiv {\rho}^{\rm out}_{0n,0n} {\rho}^{\rm 
  out}_{n0, n0} - |{\rho}^{\rm out}_{00, nn}|^2  $ is
negative. Proceeding as in the case of the attenuator, and
remembering the connection between $x_j$'s and the corresponding 
$x_j^{\,'}$'s, we have $\delta_4(\kappa,a) = ((x_1^{\,'}
x_2^{\,'} + x_3^{\,'}x_4^{\,'})/2)^2 - (|x_5^{\,'}|^2/2)^2 $. 
 The curve denoted $P_5$ in Fig.\,\ref{fig4} represents $a_4(\kappa)$
forming solution to $\delta_4(\kappa,a)=0$,  for the case of the PNES 
$(|00 \rangle
+ | 55\rangle )/\sqrt{2}$. The lines
$g_{\infty}$ and $g_1$ have the same meaning as in Fig.\,\ref{fig3}. 
The  
region $R$ between $g_{\infty} $ and $P_5$ signifies the robustness of
our PNES\,: {\em for every $\kappa \geq 1$,
  the} PNES {\em is 
seen to 
endure more noise than Gaussian states with arbitrarily large
entanglement.}     

We conclude with a pair of remarks inspired by insightful comments by the Referee.
 First, our conclusion following Eq.\,\eqref{eq3} and Eq.\,\eqref{eq10} that entanglement of two-mode 
squeezed (pure) state $|\psi(\mu)\rangle$ does not survive, for any value of $\mu$, 
channels $(\kappa,\,a)$ which satisfy the inequality $|1-\kappa^2| +a \geq 1$ applies 
to {\em all} Gaussian states. Indeed, for an arbitrary (pure or mixed) two-mode 
Gaussian state with variance matrix $V_G$ it is clear from
Eqs.\,\eqref{eq3},\,\eqref{eq10} that the  output Gaussian 
state has variance matrix 
$V^{\rm out} =\kappa^2\,V_G + (|1-\kappa^2| + a)1\!\!1_4$. Thus  
$|1-\kappa^2| + a \geq 1$ immediately implies, in view of nonnegativity of $V_G$, that  
$V^{\rm out} \geq 1\!\!1_4$, demonstrating separability of the output 
state for arbitrary Gaussian input\,\cite{simon00}. 

Secondly, Gaussian entanglement resides entirely `in' the variance matrix, and hence 
disappears when environmental noise raises the variance matrix above the vacuum or 
quantum noise limit. That our chosen states survive these environments 
shows that their entanglement resides in the higher moments, in turn
demonstrating that their entanglement is genuine non-Gaussian. Indeed,
the variance matrix of our PNES and NOON states for $N=5$ is six times
that of the vacuum state.\\
\noindent
{\em Acknowledgement\,}: K. K. Sabapathy would like to thank B. Neethi Simon for
helping with the graphs.

\end{document}